\documentclass[conference]{IEEEtran}
\makeatletter
\def\ps@headings{%
\def\@oddhead{\mbox{}\scriptsize\rightmark \hfil \thepage}%
\def\@evenhead{\scriptsize\thepage \hfil \leftmark\mbox{}}%
\def\@oddfoot{}%
\def\@evenfoot{}}

\usepackage{amsmath,amssymb,mathrsfs,amsthm,bm}
\usepackage{mathtools}
\usepackage{cite}

\usepackage{lipsum}
\usepackage{graphicx,stfloats,subfigure,multicol}
\usepackage{epsfig,epsf,psfrag,amssymb,amsfonts,latexsym,graphicx,mathrsfs,subfigure}
\usepackage[table,xcdraw]{xcolor}
\usepackage{lettrine}
\usepackage{comment}
\usepackage{enumitem}
\newlist{enumsteps}{enumerate}{2}
\setlist[enumsteps,1]{label=Case \arabic*: }
\setlist[enumsteps,2]{label=Case \arabic{enumstepsi}.\arabic*: }
\usepackage{booktabs}
\usepackage[table]{xcolor}
\usepackage{multirow}
\usepackage{tablefootnote}
\usepackage{bbm}
\usepackage{orcidlink}
\usepackage{chngpage}
\usepackage{textcomp}
\usepackage{hyperref}
\usepackage{pifont}
\usepackage{url}
\usepackage[hyphenbreaks]{breakurl}
\usepackage[normalem]{ulem}
\usepackage{algpseudocode}
\newsavebox{\ieeealgbox}

\usepackage{array}

\newtheorem*{policy*}{Dynamic NEM}
\newtheorem*{policy1*}{Generalized Dynamic NEM}

 \def\old#1{}

\def\nn{\nonumber}
\def\beq{\begin{equation}}
\def\eeq{\end{equation}}
\def\bea{\begin{eqnarray}}
\def\eea{\end{eqnarray}}
\def\ba{\begin{array}}
\def\ea{\end{array}}

\def\bitem{\begin{itemize}}
\def\eitem{\end{itemize}}
\def\ben{\begin{enumerate}}
\def\een{\end{enumerate}}

\def\ie{{\it i.e.,\ \/}}

\definecolor{bgrd}{rgb}{1,1,1}
\definecolor{gray}{rgb}{0.5,0.5,0.5}
\definecolor{dkr}{rgb}{0.7,0.1,0.2}
\definecolor{dkb}{rgb}{0.1,0.1,0.8}

\def\tcb{\textcolor{blue}}

\newcommand{\mbbR}{\mathbb{R}}

\def\Wc{{\cal W}}

\begin{document}

\title{Watts and Drops: Co-Scheduling Power and Water in Desalination Plants
}

\author{Ahmed S. Alahmed$^*$\orcidlink{0000-0002-4715-4379},
Audun~Botterud\orcidlink{0000-0002-5706-5636},
Saurabh~Amin\orcidlink{0000-0003-1554-015X}, Ali T. Al-Awami\orcidlink{0000-0003-0062-2013}
\thanks{\scriptsize  Ahmed S. Alahmed and Ali T. Al-Awami ({\tt \{\tcb{alahmad,aliawami}\}\tcb{@kfupm.edu.sa}}) are with the Electrical Engineering Department and the Interdisciplinary Research Center for Smart Mobility and Logistics at King Fahd University of Petroleum and Minerals, Dhahran, Saudi Arabia. Audun Botterud and Saurabh Amin ({\tt \{\tcb{audunb,amins}\}\tcb{@mit.edu}}) are with the Laboratory for Information and Decision Systems at  Massachusetts Institute of Technology, Cambridge, MA, USA. ({$^*$\em Corresponding author: Ahmed S. Alahmed.})}
\thanks{\scriptsize This work was supported in part by the Interdisciplinary Research Center for Smart Mobility and Logistics at King Fahd University of Petroleum and Minerals, under project no. INML2411.}
}

\maketitle
\begin{abstract}
We develop a mathematical framework to jointly schedule water and electricity in a profit-maximizing, renewable-colocated water desalination plant that integrates both thermal- and membrane-based technologies. The price-taking desalination plant sells desalinated water to a water utility at a given price and engages in bi-directional electricity transactions with the grid, purchasing or selling power based on its net electricity demand. We show that the optimal scheduling policy depends on the plant’s internal renewable generation and follows a simple threshold structure. Under the optimal policy, thermal-based water output decreases monotonically with renewable ouput, while membrane-based water output increases monotonically. We characterize the structure and intuition behind the threshold policy and examine key special properties.
\end{abstract}


\section{Introduction}\label{sec:intro}
The water-energy nexus has garnered growing research attention. The interdependence is bidirectional: electricity powers water extraction, treatment, and desalination, while water is essential for power generation, especially in thermal plants. This mutual reliance has prompted the development of integrated models to optimize resource use and support cross-sectoral planning. Thermal desalination plants (TDPs), commonly used in energy-rich regions, produce freshwater as a primary product while producing electric power as a byproduct. These cogeneration plants benefit from shared heat and power flows and result in tightly coupled operational dynamics. Notably, \cite{Santhosh&Farid&Toumi:14AE,ToumiEnergy2014} developed dynamic physics-based models for integrated thermal power and thermal water desalination systems, highlighting the importance of coordinated management of thermal power plants and thermal desalination plants.

In contrast, reverse osmosis (RO) has recently become the leading desalination technology due to its modularity and improved energy efficiency. In fact, according to some reports, RO now accounts for approximately 69\% of the installed desalination capacity globally, with most new projects deploying RO desalination units \cite{Eke2020}. RO desalination plants (RODPs) primarily draw electricity from the grid, positioning them as large, flexible loads within the power system. Recent studies in \cite{Ghaithan2022,Schar2023,Qudah2024} have explored coupling RODPs with renewable energy sources using hybrid optimization frameworks to minimize costs and emissions. However, these studies treat the power system as an external input and overlook the bidirectional interactions between water and energy. This gap is addressed in \cite{MoazeniAPEN2020,MoazeniJCP2020,AlAwamiTPS2022} by modeling co-optimized water-energy microgrids and coordinated RO-power scheduling platforms, considering dynamic demand, grid constraints, and renewable variability. Their research also explores the use of RODPs for demand-side grid support, including demand response and frequency regulation. 

 However, the existing literature has not addressed the interaction between thermal and RO desalination units when they co-exist within the same territory in the water-energy nexus context.

This paper develops a mathematical framework for the joint scheduling of water and power in a {\em profit-maximizing} water desalination plant (WDP) equipped with thermal and membrane-based technologies, and local renewable generation. The WDP is modeled as a price-taking agent that engages in water and electricity transactions with a water utility and an electric utility. A key challenge lies in utilizing the complementarities between the two different desalination technologies, especially when coupled with the variability of the renewable energy supply.

Our main contribution is the characterization of WDP's optimal production plan when colocated with renewables. We show that the optimal policy adopts a simple two-threshold structure: when renewable generation is low, the WDP prioritizes thermal-based desalination, and as renewable availability increases, it shifts progressively toward membrane-based production. This structural result provides not only a computationally tractable policy but also valuable operational insights. Specifically, we prove that the thermal-based water output decreases monotonically with renewable availability, while membrane-based output increases accordingly.

In addition to the general solution structure, we analyze several special cases to illustrate the practical implications of the proposed policy and to highlight conditions under which either technology is preferred. Our findings have broader relevance for the optimal dispatch of hybrid energy-water systems, particularly in smart grid environments where flexibility and responsiveness are increasingly rewarded.

\section{WDP Framework and Decision Problem}\label{sec:form}
\begin{figure}
    \centering
    \includegraphics[scale=0.67]{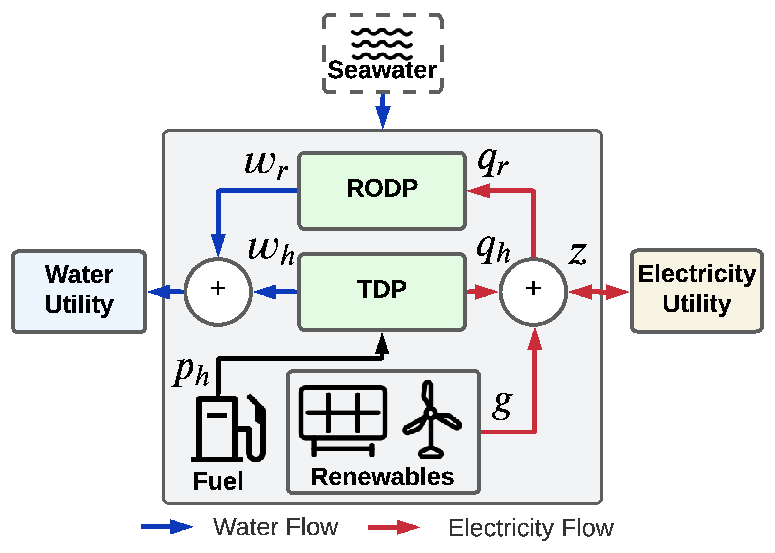}
    \caption{Hybrid WDP framework with TDP, RODP, and renewables. The water outputs of the thermal and membrane units are denoted by $w_h, w_r \in \mbbR_+$, respectively. The electricity output of the TDP, RODP, and renewables are denoted by $q_h, q_r,g \in \mbbR_+$, respectively. The net electricity consumption of the WDP is denoted by $z \in \mbbR $. The amount of fuel feeding the TDP is denoted by $p_h \in \mbbR_+$.}
    \label{fig:WDP}
\end{figure}

We consider a hybrid WDP that jointly optimizes water and electricity schedules to maximize profit, defined as the revenue from selling water and electricity minus operational costs, including electricity procurement from the grid (Fig.~\ref{fig:WDP}). The WDP utilizes thermal and membrane-based technologies, in addition to renewable energy sources.\footnote{Although membrane-based technologies are improving rapidly, thermal desalination remains more robust under high-salinity and harsh climatic conditions, while also offering better grid support capabilities.}
On one hand, the WDP unidirectionally interacts with the water utility by selling water at a price that is known {\em apriori} and may be temporally varying. On the other hand, the WDP bidirectionally transacts with the electricity utility by buying and selling electricity at possibly asymmetric prices of buying and selling based on its net consumption. In both markets, water and electricity, the WDP operates as a {\em price-taker}.

The WDP electric grid net metering arrangement here is similar to that of energy communities and distributed energy resources aggregation schemes under regulated utilities \cite{Alahmed&Tong:24TEMPR,Chakraborty&Poolla&Varaiya:19TSG}.

For the remainder of this section, we present the modeling of the TDP, RODP, payment functions of the water and electricity utilities, and the WDP co-optimization problem.

\subsection{TDP and RODP Models}

\subsubsection{TDP}
The TDP uses a boiler to heat seawater using a fuel source, producing steam that is then condensed to obtain fresh water. The waste heat is used to power a turbine to produce electricity. The TDP, therefore, produces both water and electricity, which we denote by $w_h \in \mbbR_+$ and $q_h \in \mbbR_+$, respectively. The water generated by the TDP is a function of burned fuel $w_h = f_h(p_h) = \alpha_h p_h$, where $\alpha_h \geq 0$ is a thermal conversion factor. Let $f'_h(p_h)$ be the derivative of $f_h(p_h)$ with respect to $p_h$.

The TDP produces electricity as a by-product of water desalination, governed by the relation $w_h = \eta_h q_h$, where $\eta_h > 0$ denotes the water-to-electricity production ratio. We further define the thermal conversion factor as $\beta_h := \alpha_h / \eta_h \geq 0$, representing the ratio of fuel energy to electricity output.

The TDP water flowrate is bounded by the maximum $\overline{w}_h \in \mbbR_+$ and minimum $\underline{w}_h \in \mbbR_+$ operational limits that ensure stable operation, \ie $w_h \in [\underline{w}_h, \overline{w}_h]$.

The TDP's operating cost function is denoted by $C_h(p_h)\in \mbbR$. We assume the cost function $C_h(\cdot)$ to be strictly convex, continuously differentiable, and non-decreasing.

\subsubsection{RODP}
The RODP consumes electricity to force pressure seawater through a semi-permeable membrane under high pressure, effectively filtering out salt and impurities.

The water production and electricity consumption of the RODP are denoted by $w_r, q_r \in \mbbR_+$, respectively. The amount of water produced by the RODP $w_r$ is a linear function of the electricity consumed $q_r$ by that module, \ie $w_r = f_r(q_r) = \alpha_r q_r$, where $\alpha_r> 0$ is the RODP's conversion factor. We use $f_r'$ to denote the derivative of $f_r$ with respect to $q_r$.

The water production of the RODP is bounded from above by the maximum $\overline{w}_r \in \mbbR_+$ flowrate  and from below by the minimum $\underline{w}_r \in \mbbR_+$ operational limits that ensure stable operation, \ie $w_r \in [\underline{w}_r, \overline{w}_r]$.

\subsection{Water and Electricity Payments}

 \subsubsection{Water Payment}
The profit-maximizing WDP sells the total desalinated water from the TDP and RODP to the water utility; therefore, the WDP revenue function is given by
\begin{equation}\label{eq:WaterPayment}
    P^w(w_h,w_r) = \pi^w ( w_h +  w_r),
\end{equation}
where $\pi^w \in \mbbR_+$ is the water selling price. 

\subsubsection{Electricity Payment}
The profit-maximizing WDP transacts with the electric utility, which charges the WDP based on its net electricity consumption $z: \mbbR_+\times \mbbR_+ \rightarrow \mbbR$, given by
\begin{equation}\label{eq:NetConsumption}
    z(q_r,q_h;g) =  q_r-  q_h - g,
\end{equation}
where $g \in \mbbR_+$ is the WDP aggregate renewables production.

Given the electric utility's prices, the electricity payment function is given as
\begin{equation}\label{eq:ElectricPayment}
    P_\pi(q_r,q_h;g)=\pi^+[z(q_r,q_h;g)]^+ -\pi^-[z(q_r,q_h;g)]^-,
\end{equation}
where $[x]^+, [x]^-$ are the positive and negative parts of $x$, \ie $[x]^+=\max\{0,x\}$, $[x]^- =-\min\{0,x\}$, and $x= [x]^+ - [x]^-$. The electricity tariff parameters $(\pi^+,\pi^-)\in \mbbR_+$ are the power import and export prices, respectively.  To avoid utility death spirals and risk-free arbitrage by the WDP, we assume $\pi^+ \geq \pi^-$ \cite{Alahmed&Tong&Zhao:24TSE}.

The WDP imports power if $z>0$, and exports power if $z<0$.

\subsection{WDP Profit}
The profit function of the profit-seeking WDP can be written as
\begin{align}\label{eq:WDPprofit}
    \Pi(w_h,w_r,q_h,q_r;g):=& P^{w}(w_h,w_r) - P_\pi(q_r,q_h;g)\nn\\& -  C_h(p_h),
\end{align}
where the first term represents the revenue generated from water sales; the second represents the monetary transaction from net electricity consumption, constituting a cost when electricity is purchased ($z>0$) and a revenue when it is sold ($z<0$); and the third term reflects the operating costs of the TDP. The operating costs of the RODP, by contrast, are implicitly accounted for through the cost of procuring the energy $ q_r$, which may be sourced from on-site renewables, the TDP, the grid, or any combination thereof.

\subsection{WDP Decision Problem}
The WDP’s objective is to maximize its profit subject to plant capacity and energy balance constraints by co-optimizing the thermal and RO units; hence, it solves the following program
\begin{align}\label{eq:Optimization}
 (w_h^\ast,w_r^\ast,q_h^\ast,q_r^\ast) :=\hspace{-0.3cm}&\underset{w_h,q_h ,w_r,q_r \in \mbbR_+}{\operatorname{argmax}} \hspace{0cm}\Pi(w_h,w_r,q_h,q_r;g)\nn\\ 
 	&~~~~~~\text{subject to}~~~ z=  q_r - q_h - g\nn\\
    &\hspace{2.5cm} w_h + w_r \geq \Wc  \nn\\
		&\hspace{2.5cm}	 \underline{w}_h \leq w_h \leq \overline{w}_h\nn\\
        &\hspace{2.5cm}	 \underline{w}_r \leq w_r \leq \overline{w}_r,
\end{align}
where $\Wc$ is the total desalinated water demanded by the water utility. Note that the objective of the profit-maximization problem above is non-differentiable due to the presence of indicator functions in the electricity payment expression in (\ref{eq:ElectricPayment}). We assume that the WDP is sized so that the minimum output of desalinated water is no less than that demanded by the water utility, \ie $\underline{w}_h + \underline{w}_r \geq \Wc$.


\section{Optimal WDP Operation}\label{sec:Optimal}
We highlight here the optimal co-scheduling policy and its properties, leaving detailed theoretical development in \cite{Alahmed&Botterud&Amin&Alawami:25WorkingPaper}.

\begin{figure}
    \centering
    \includegraphics[scale=0.55]{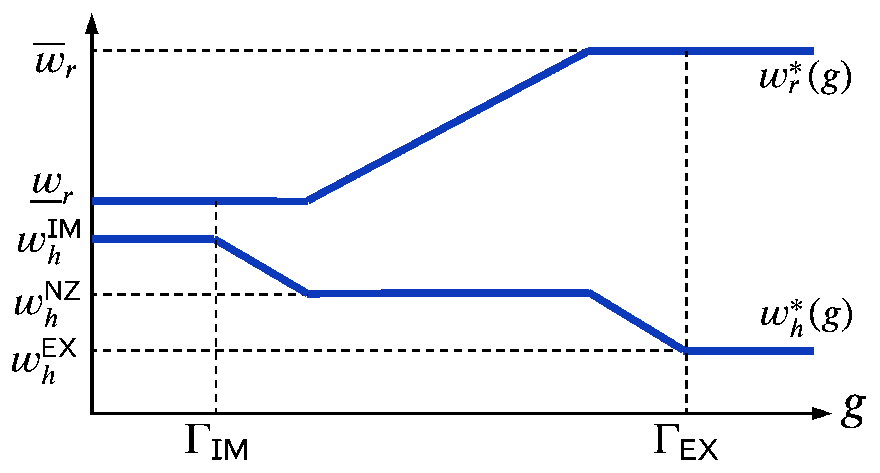}
    \caption{Optimal TDP and RODP water schedules, when $f_r' \pi^w \in [\pi^-, \pi^+]$ and $w^{\sf{IM}}_h<w^{\sf{IM}}_r$.}
    \label{fig:Optimal}
\end{figure}

The optimal threshold-based co-scheduling policy for TDP and RODP when $f_r' \pi^w \in[\pi^-, \pi^+]$ is depicted in Fig. \ref{fig:Optimal}. To schedule the plants, the WDP measures the aggregate renewable generation $g$ and compares it with two thresholds ($\Gamma_{\sf{IM}}, \Gamma_{\sf{EX}}, \Gamma_{\sf{EX}}\geq \Gamma_{\sf{IM}}$). The WDP imports power ($z>0$) for water desalination when $g<\Gamma_{\sf{IM}}$, exports power ($z<0$) and water when $g>\Gamma_{\sf{EX}}$, and operates in an islanded-mode with respect to the electricity utility ($z=0$) when $g \in [\Gamma_{\sf{IM}},\Gamma_{\sf{EX}}]$, while using all local generation to desalinate and sell water. Therefore, when $g<\Gamma_{\sf{IM}}$, the WDP uses local renewables, the TDP-produced electricity, and the grid electricity to feed the RODP's consumption of electricity to produce water. When the renewables are at a higher level of $g \geq \Gamma_{\sf{IM}}$, the WDP uses only local renewables and TDP-produced electricity for the RODP's consumption.

The two thresholds that govern the optimal policy, $\Gamma_{\sf{IM}}$ and $\Gamma_{\sf{EX}}$, depend only on the parameters of water and electricity tariffs and the conversion coefficients of the TDP and RODP. Both $\Gamma_{\sf{IM}}$ and $\Gamma_{\sf{EX}}$ can be computed {\em offline}, independent of the renewable generation $g$.

As depicted in Fig.~\ref{fig:Optimal}, the optimal TDP water output decreases monotonically with $g$, dynamically transitioning from the three constant production levels $w^{\sf{IM}}_r$, $w^{\sf{NZ}}_h$, and $w^{\sf{EX}}_r$. The TDP outputs are all computed in closed form. On the other hand, the optimal RODP water output increases monotonically with $g$, dynamically transitioning from the minimum production level to the maximum production level. The RODP thresholds are also computed in closed-form, and independent of the electric tariff parameters \cite{Alahmed&Botterud&Amin&Alawami:25WorkingPaper}.

It should be noted that the WDP's profit is linear in $w_r$, yielding boundary solutions when $f_r' \pi^w > \pi^+$, with $w_r^\ast(g) = \overline{w}_r$ for all $g$, and when $f_r' \pi^w < \pi^-$ with $w_r^\ast(g) = \underline{w}_r$ for all $g$. An implication of this linearity is that if $f_r' \pi^w>\pi^-$, the TDP's electricity output is never exported, unless RODP reaches its upper operational limit $\overline{w}_r$. Also, under this case, the WDP prioritizes using renewable energy for water desalination rather than energy export. When $f_r' \pi^w > \pi^+$, energy exports only occur if RODP saturates. The intuition here is that using renewables to desalinate and sell water is always more lucrative to the WDP than selling the renewables directly to the electric utility.

We analyze in \cite{Alahmed&Botterud&Amin&Alawami:25WorkingPaper} some special cases, and comparative statics of the optimal decisions with changing tariff and desalination technology parameters.

    \section{Numerical Results}\label{sec:num}
To showcase the optimal operation of the plant and the effect of tariff and plant parameters, we consider a WDP with TDP, RODP, and solar PV renewable generation. Semi-synthetic data are used in the simulation. The WDP faces a {\em water utility} that purchases desalinated water at a price $\pi^w$, which is varied throughout the simulation. The WDP also faces an {\em electric utility} that adopts a net energy metering tariff with a retail rate of $\pi^+=\$0.4$/kWh and a sell rate of $\pi^- = \$0.1$/kWh, respectively.

The TDP operating cost function is assumed to be a convex quadratic function 
$   C_h(p_h)= b p_h^2 + a p_h$, with $b=\$0.001/\text{BTU}^2$ and $a=\$0.05/\text{BTU}$. The TDP and RODP flowrate limits are set as follows $\overline{w}_h=\overline{w}_r=8333\: m^3h, \underline{w}_h=\underline{w}_r=0 \: m^3h$. The conversion factor of TDP is $\alpha_h= \eta_h= 1/10$, whereas the conversion for RODP is set to $\alpha_r= 1/4$.

\begin{figure}[htbp]
    \centering
    \includegraphics[width=0.7\linewidth]{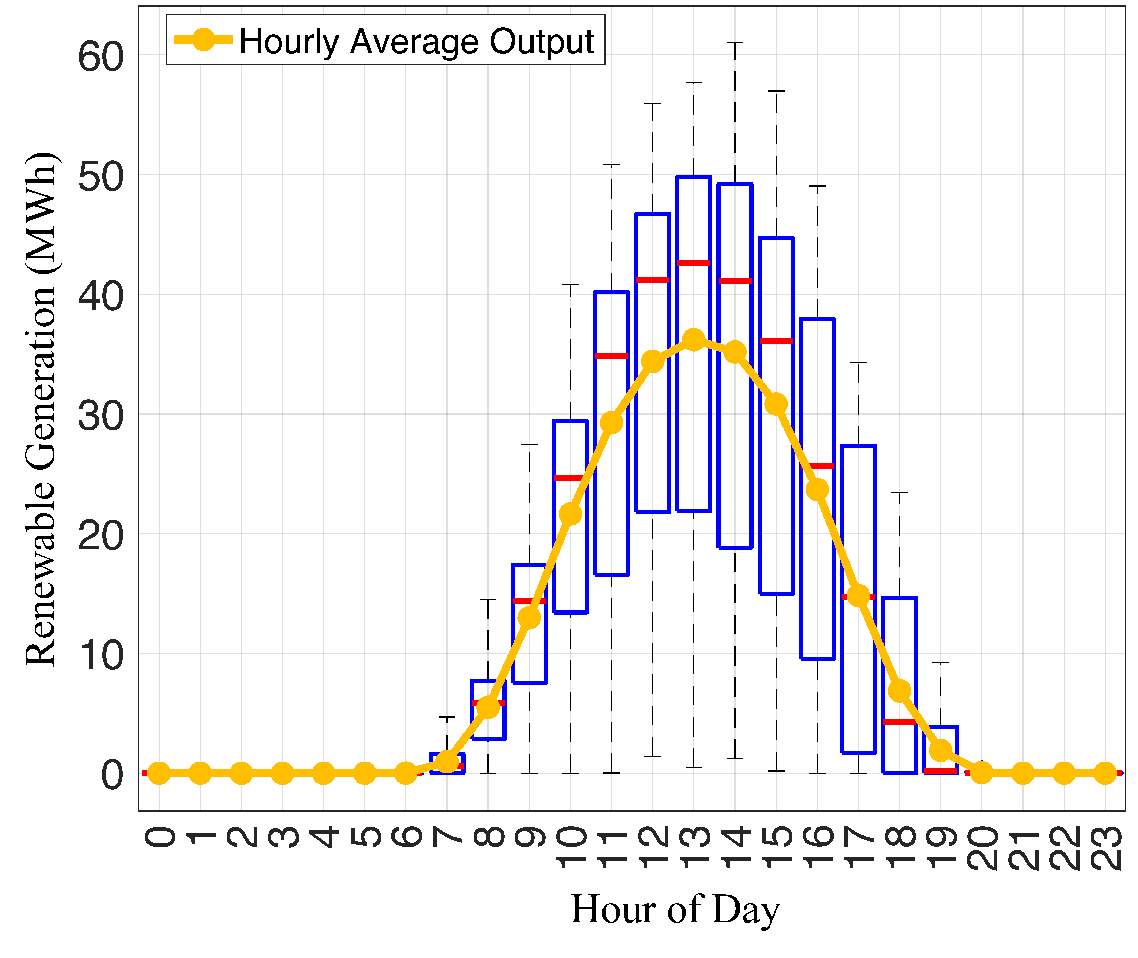}
    \caption{Daily Solar PV generation profile.}
    \label{fig:Rawdata}
\end{figure}

We used PecanStreet's 2018 rooftop solar data \cite{PecanStreet} by aggregating the generation output of 23 households and then scaling it up to meet WDP's energy needs. The distribution of the annual aggregate and scaled renewable generation profile is shown in Fig.\ref{fig:Rawdata}. To capture variability in renewable output, we performed 100,000 Monte Carlo simulations using hourly profiles generated as independent normal random variables with hour-specific means and variances derived from historical data.

On average, solar photovoltaic output peaked in the afternoon at 13, reaching approximately 36.2 MWh, with mean outputs of 34.4 and 35.2 at hours 12 and 14, respectively. PV output is negligible, recording zero from 1 to 6 and 21 to 23, and remained near zero at 7 and 20. During high production hours, the median (red line) was higher than the mean, indicating a negatively skewed distribution. This skewness is attributed to the relatively high production on most days, contrasted by notably low production during heavily overcast days, which created longer tails on the left side of the distribution. 

\subsection{Optimal Water and Electricity Schedules}
We examine, in Figures \ref{fig:Num2}-\ref{fig:Num1}, the optimal water and energy scheduling decisions of the WDP under three water price scenarios, with the intermediate case being the most interesting. Across all three figures, blue and red curves denote the RODP and TDP outputs, respectively, while green and yellow curves denote net electricity consumption and renewable energy production, respectively.

\subsubsection{High Water Price}
\begin{figure}[htbp]
    \centering
    \includegraphics[width=1.0\linewidth]{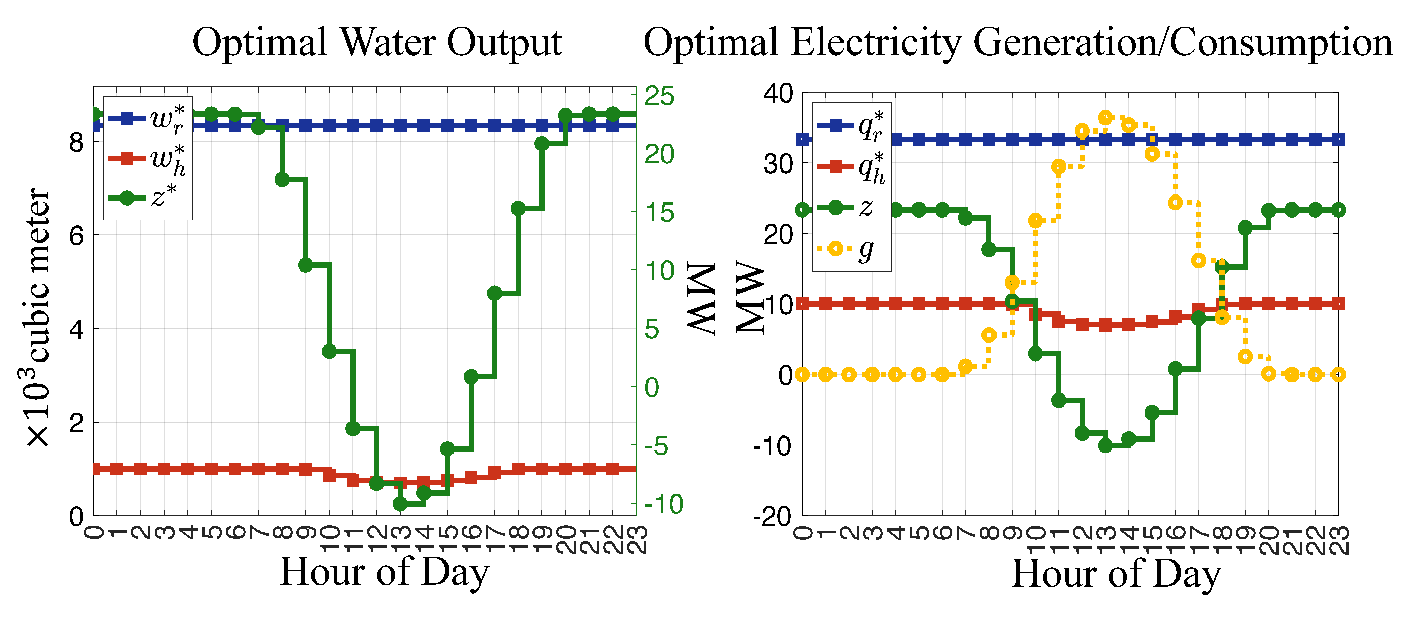}
    \caption{WDP's optimal water and electricity profiles under $\pi^w = \$5/m^3h$.}
    \label{fig:Num2}
\end{figure}

When the price of water was set at $\pi^w = \$5/m^3h$, the revenue from selling desalinated water to the water utility was higher than the cost of purchasing electricity from the electric utility, therefore higher than the revenue from selling electricity, the WDP scheduled RODP water at maximum capacity by buying from the grid, consuming the TDP generated power and consuming the local renewables (Fig.\ref{fig:Num2}). Therefore, as shown in Fig.\ref{fig:Num2}, the RODP output was always set at the maximum, \ie $w^\ast_r = \overline{w}_r=8,333\:m^3h$.

The TDP was scheduled at the marginal price of the WDP, which is the sum of the grid price and the effective price of selling water, \ie, after accounting for the conversion factor, resulting in an output of around 1,000 $m^3h$. A kink in the TDP output can be seen when the plant shifted from net-consuming to net-producing, causing the marginal price to change from $f'_h \pi^w+\pi^+$ to $f'_h \pi^w+\pi^-$. The intuition of the kink can be better explained using the electricity plot on the right. The increasing renewable output replaced the need for grid purchases, which caused the net electricity consumption to drop, as can be seen in hours 7 to 13. Having higher renewables reduced the need for TDP-based power generation to feed the RODP. The plant became a power exporter during hours 11 to 15, \ie negative net electricity consumption, when the renewable power together with the reduced TDP-based power were more than enough to supply the RODP, which led to excess electricity being sold to the grid.

\subsubsection{Low Water Price}
\begin{figure}[htbp]
    \centering
    \includegraphics[width=1.0\linewidth]{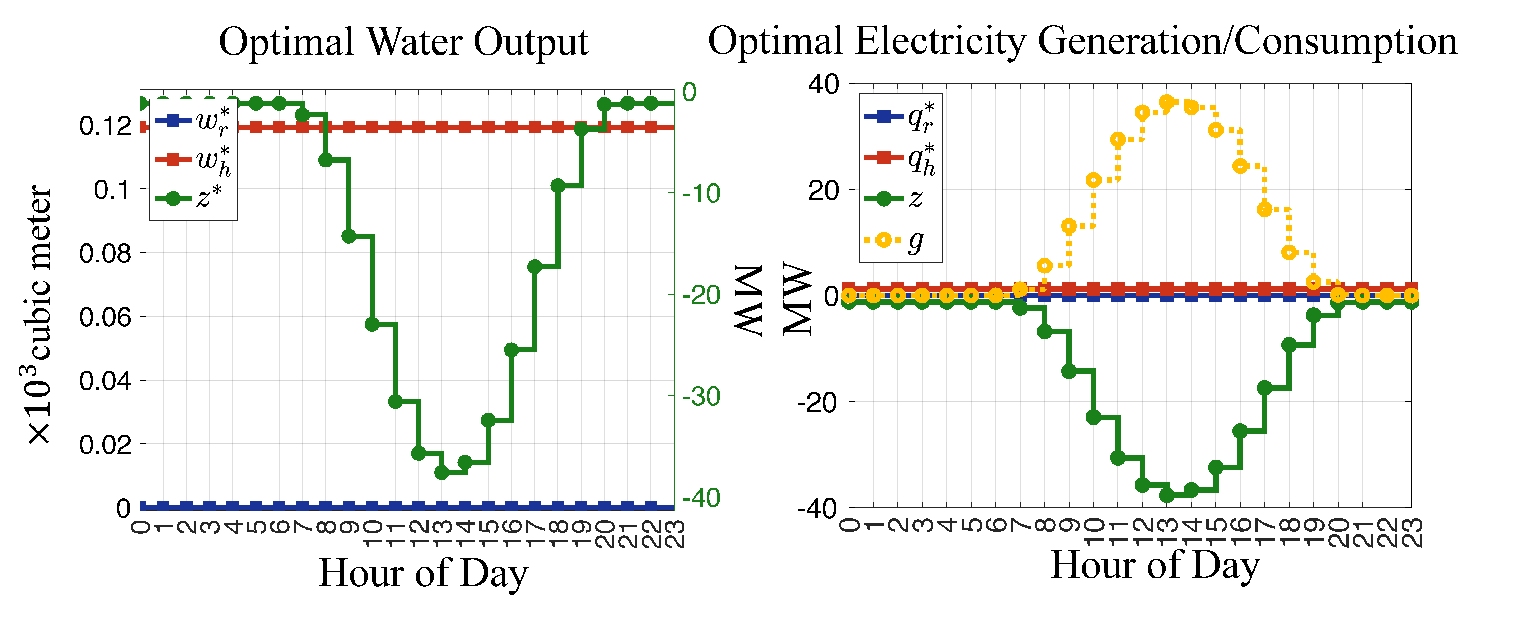}
    \caption{WDP's optimal water and electricity profiles under $\pi^w = \$0.2/m^3h$.}
    \label{fig:Num3}
\end{figure}

As shown in (Fig.\ref{fig:Num3}), here, we set the price at $\pi^w = \$0.2/m^3 h$, which resulted in scheduling the RODP at the minimum $w^\ast_r = \underline{w}_r=0\:m^3 h$. The reason is that under such $\pi^w$, the revenue from desalinating and selling water is less than the cost of buying the energy needed from the grid, and less than the revenue from selling electricity to the electric utility; therefore, the plant only sells water when it comes as a by-product of generating electricity, \ie from the TDP, which generated 1,093 $m^3 h$ of water and 1.093 MWh of electricity. The TDP's water and electricity generation were constant, because the marginal price of the plant was constant throughout the day as the WDP was always in the net-production zone, \ie negative net electricity consumption. All generated renewables were sold back to the grid, as such action generates more revenue for the plant than using the electricity to desalinate and sell water.

\subsubsection{Intermediate Water Price}
\begin{figure}[htbp]
    \centering
    \includegraphics[width=1\linewidth]{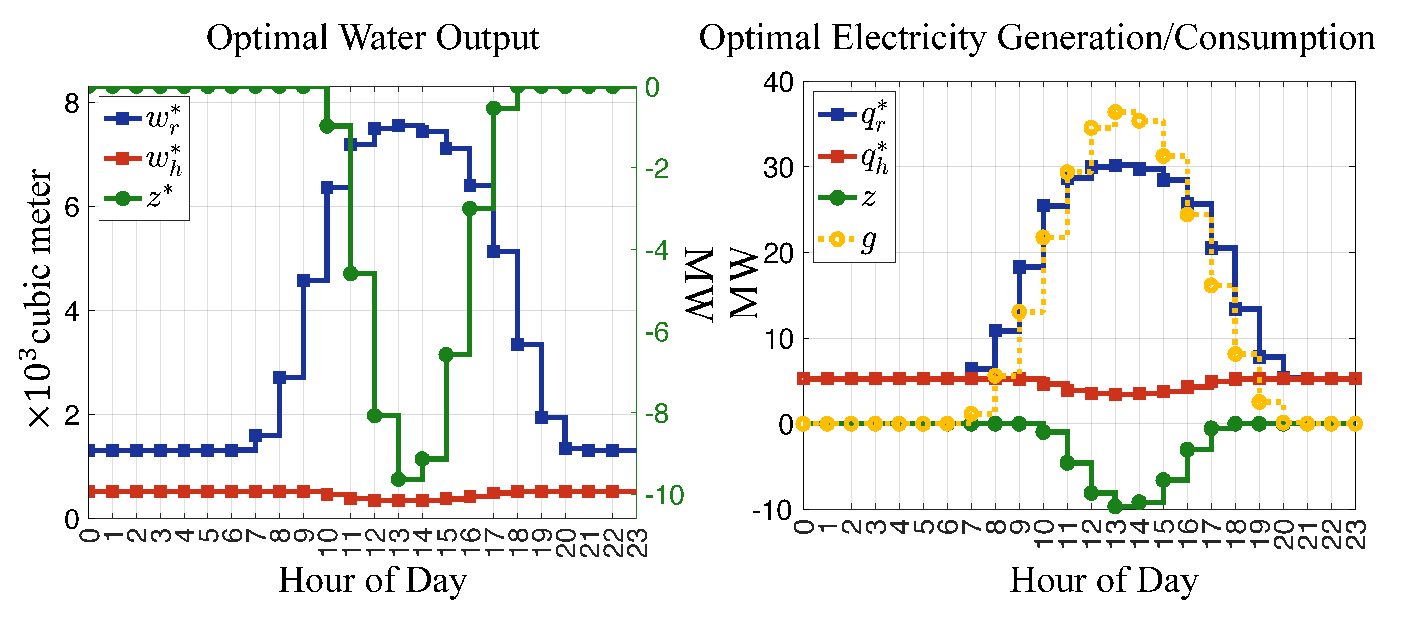}
    \caption{WDP's optimal water and electricity profiles under $\pi^w = \$1.5/m^3h$.}
    \label{fig:Num1}
\end{figure}

When the price of water was set at $\pi^w = \$1.5/m^3 h$, the revenue from selling desalinated water became higher than the revenue from selling the electricity that can be used to desalinate water, but less than the cost of buying electricity from the grid to desalinate water. As shown by the green curve in (Fig.\ref{fig:Num1}), the WDP was either a net zero or a net electricity producer. As shown in the right panel of Fig.\ref{fig:Num1}, when the renewable output was zero, the RODP only consumed the TDP-based generated power of 5.2 MWh, \ie without consuming from the grid. In hours 7-9 and 18-19, the renewable output was fully consumed by the RODP, in addition to TDP's power output, keeping the WDP in the net-zero zone, \ie off the electric grid. When the renewable generation was higher than what the RODP can consume, e.g., hours 10-17, the excess renewable was sold as electricity to the electric grid at $\pi^-$, causing the marginal price of the plant to shift, which subsequently resulted in scaling down the production of the TDP, due to the abundance of renewables. The WDP co-optimization in this scenario is highly dependent on the renewable generation output, with the RODP consumption becoming a follower of the TDP and renewable power generation.

\section{Conclusion}\label{sec:conclusion}
 We analyzed the optimal scheduling decisions of a profit-maximizing WDP with thermal and RO-based desalination technologies that face water and electric utilities. We showed that the WDP is optimally scheduled based on a two-threshold policy based on the renewable generation output. As the output of renewable generation increases, the WDP monotonically decreases the water flow of the TDP and increases that of the RODP. This structural result provides not only a computationally tractable solution but also valuable operational insights. Specifically, we show that when the revenue from selling desalinated water is higher than the price of selling electricity, the TDP and renewables electricity outputs are fully consumed locally by the RODP to produce and sell more water. The RODP output is always set to maximum if the revenue from selling water is higher than the cost of buying electricity. It is set to a minimum if the revenue from selling water is less than the revenue from selling electricity.

{
\bibliographystyle{IEEEtran}
\bibliography{WattsDrops}
}


\end{document}